\documentclass[twocolumn]{aastex631}

\def\msun{\,M_{\odot}}

\usepackage{amsmath}
\usepackage{bm}
\usepackage{float}
\usepackage{graphicx}
\usepackage{booktabs}
\usepackage[toc,page]{appendix}

\begin{document}

\title{Hidden Twin Star Solutions from an Agnostic Speed-of-Sound Model: Confronting XTE J1814--338's Extreme Compactness}

\author[0009-0000-8504-9134]{Tianzhe Zhou}
\affiliation{Department of Physics, Tsinghua University, Beijing 100084, China}
\email{zhoutz22@mails.tsinghua.edu.cn}
\author[0000-0001-6406-1003]{Chun Huang}
\affiliation{Physics Department and McDonnell Center for the Space Sciences, Washington University in St. Louis; MO, 63130, USA;\\}
\email{chun.h@wustl.edu}

\begin{abstract}
The twin star configuration, where two neutron stars share the same mass but exhibit different radii, arises from a strong first-order phase transition within the stellar interior. In widely used equation of state (EoS) meta-models, such as the Polytrope (PP) and Speed-of-Sound (CS) models, this first-order phase transition behavior can be naturally mimicked by tuning some model parameters. Here, we systematically explore the under-explored parameter space within one of a widely adopted CS model that leads to twin stars via a strong first-order phase transition. Within this twin-star subspace, we perform a comprehensive Bayesian analysis that integrates mass--radius (MR) constraints from X-ray observations of rotation-powered millisecond pulsars. The resultant twin star branch, situated within the 1--1.2 $M_{\odot}$ mass range and approximately 7 km in radius, surprisingly coincides with the MR ranges proposed for the recent anomaly in the Accreting Millisecond X-ray Pulsars XTE J1814--338 (J1814), suggesting a hybrid twin star configuration. Moreover, incorporating the J1814 observation as an additional constraint yields an extreme phase transition pressure $P_{\text{trans}} = 108.9_{-4.85}^{+6.46}$ MeV/fm$^3$, a transition density of $\varepsilon_{\text{trans}}/\varepsilon_0 = 4.847_{-0.134}^{+0.271}$(where $\varepsilon_0$ is the nuclear saturation energy density) and an energy density jump $\Delta \varepsilon = 558.7_{-278.7}^{+303.6}$ MeV/fm$^3$, corresponding to $\Delta \varepsilon/\varepsilon_0 = 3.716_{-1.854}^{+2.020}$. Notably, to satisfy all astrophysical constraints, the speed of sound inside of the hybrid twin star core is driven toward the speed of light ($c_s^2/c^2 > 0.9$), indicating the potential presence of strongly interacting, exotic matter in this core region.
\end{abstract}


\section{Introduction} \label{sec:intro}
Widely used dense matter equation-of-state (EoS) meta-models—such as the speed-of-sound (CS) model \citep{Tews_2018,10.1093/mnras/stz654} and the polytropic model \citep{Hebeler_2013}—are designed to span the broadest possible mass–radius (MR) parameter space and to replicate the full spectrum of EoS behaviors. Numerous studies have exploited these flexible frameworks to investigate the constraints imposed by multi-messenger observations and nuclear experiments (e.g., \cite{Tews:2018kmu,Raaijmakers_2019,Raaijmakers_2020,Raaijmakers_2021,Rutherford_2024}), while others have focused on physics-motivated phenomenological EoS (e.g., \cite{Traversi_2020,Malik_2022,Huang:2023grj,Huang:2024rvj,2025arXiv250115810L}). By adjusting the model parameters, a wide variety of EoS behaviors can be simulated. For instance, a first-order phase transition can be readily generated within the meta-model framework. When a strong phase transition occurs in the core of a neutron star, an alternate MR branch may emerge—yielding two stars with the same mass but different radii, a phenomenon known as `twin stars' \citep{PhysRev.172.1325,Kampfer_1981,glendenning1998nonidenticalneutronstartwins}. The existence of twin stars thus provides a valuable probe into the internal phase transitions of neutron stars, and several studies have examined how astrophysical observations and nuclear experimental constraints affect the potential for twin star formation (e.g., \cite{Montana:2018bkb,mendes2024constrainingtwinstarscold,Christian:2017jni}).

Recently, \citep{KeithNICER}, detailed X-ray pulse modeling of the accreting millisecond X-ray pulsar XTE J1814--338 using RXTE data \citep{Kini_2024} has revealed that this object exhibits an unusually small mass and radius compared to previous MR measurements. Some studies have proposed that this anomaly could be explained by a hybrid star configuration, a neutron star featuring an inner core of exotic matter (e.g., deconfined quark matter) surrounded by a conventional hadronic envelope \citep{laskospatkos2024xtej1814338potentialhybrid,Veselsky:2024bnf}, or even a dark matter admixed neutron star \citep{lopes2025xtej1814338darkmatter,Pitz:2024xvh}. In this study, we adopt an agnostic construction--widely used in NICER posterior analyses--to first explore the potential of the CS model to produce twin star configurations, and then compare the predicted twin star MR posterior with the anomalous observation of XTE J1814--338 (J1814).

While most previous works have focused on piecewise EoS constructions that manually introduce first-order phase transitions to generate twin stars (e.g., \cite{huang2025,2021PhRvD.103f3026T,PhysRevD.108.094014,PhysRevD.108.043013,Tang:2021snt,Gorda_2023,PhysRevC.107.025801,komoltsev2024firstorderphasetransitionscores,christian2025}), the CS construction inherently permits phase transitions and is sufficiently general to capture a wide range of EoS behaviors.
And in a recent survey by \cite{verma2025}, the non-monotonic parametrized EoS models, as we employed herein, exhibited the largest Bayesian evidence in accounting for current astronomical observations, outperforming both the monotonous and discontinuous models.
Rather than exploring the entire CS model parameter space, we perform a comprehensive Bayesian analysis focused on a `latent' subspace—one that is capable of producing twin stars through strong phase transitions. Despite indications of its existence (see, e.g., \cite{Rutherford_2024,huang2025}), this parameter space has received limited discussion in the literature. Here, we directly constrain this subspace using current astrophysical observations, thereby supporting the hypothesis that the twin star configuration could explain the anomalously compact MR observation of XTE J1814--338.

The remainder of the paper is organized as follows. In Section \ref{sec:method}, we describe the methodology used to explore the twin star parameter space and outline the Bayesian inference framework employed in our study. Section \ref{sec:result} presents the posterior results under conditions that permit twin stars, highlighting the parameter subspace that meets current astrophysical constraints and comparing these results with the MR measurements of XTE J1814--338. Finally, Section \ref{sec:discuss} concludes with a summary and discussion of our work.

\section{Method} \label{sec:method}

Various speed‐of‐sound formulations have been proposed to explore an adequately broad MR and EoS parameter space (e.g., \cite{Greif2019, Tews_2018}). Here, we focus on the CS model as implemented in \cite{Raaijmakers_2019, Raaijmakers_2020, Raaijmakers_2021, Rutherford_2024}, which originates from the work of \cite{Greif2019}. However, these studies lack a detailed discussion of the scenarios in which phase transitions may occur, particularly in the context of strong phase transitions and extremely high-density conditions. This represents an under-explored area that our work aims to address. We will first present our EoS setting, which is consistent with previous studies based on the same model, and then motivate the extension of this CS model to an even broader range to accommodate the existence of twin star in a reasonable mass range. Finally, the Bayesian framework utilized in this work will be described, followed by an examination of the conditions necessary for the existence of twin stars.
\subsection{EoS settings} \label{sec:EoS}
In this subsection, the equation of state (EoS) setup is briefly outlined. The neutron star EoS is divided into three regions: the outer crust, the inner crust, and the core. In this work, we employ the BPS crust with a log‐linear interpolation for the outer crust region \citep{BPS_crust}, defined for densities $\varepsilon < \varepsilon_{\text{outer}} = 4.30 \times 10^{11}$ g/cm$^3$. We then define a polytropic segment to seamlessly connect the BPS crust from a density of $\varepsilon_{\text{outer}}$ to $0.5\varepsilon_0=1.336 \times 10^{14} \, \rm g/cm^3$, representing the inner crust EoS,
where $\varepsilon_0=m_{\mathrm{N}} n_0$ is the energy saturation density, 
$n_0 = 0.16\,\mathrm{fm}^{-3}$ is the nuclear saturation density, and $m_{\mathrm{N}} = 939.565\,\mathrm{MeV}$ is the nucleon mass.
Throughout this paper, we set $c=1$ when no ambiguity arises.

Although this choice differs from the $\chi$EFT band constraint for the inner crust as outlined in \cite{Greif2019}, it avoids the complexities associated with the nuclear pasta structure in this density range, consistent with the inner crust treatment described in \cite{Huang:2023grj,Huang:2024rvj}. We have confirmed that this approach remains consistent with the $\chi$EFT band while being better motivated from a nuclear physics perspective; further details are available in \cite{Huang:2023grj}.

For the core part, we adopt a speed-of-sound parametrization as proposed in \cite{Greif2019}. The speed of sound square in neutron star matter, expressed as $c_s^2(x)/c^2$, at a given normalized density $x$ is parametrized as:
\begin{equation}
\frac{c_s^2(x)}{c^2} = a_1\, \mathrm{e}^{-\frac{1}{2}\frac{(x-a_2)^2}{a_3^2}} + a_6 + \frac{a_7 - a_6}{1+\mathrm{e}^{-a_5\,(x-a_4)}},
\label{cs_eq}
\end{equation}
where $x = \varepsilon/\varepsilon_0$.
The parameters $a_1$ through $a_7$, excluding $a_6$ , are free parameters tuned to cover a sufficiently broad region of the EoS parameter space.
In contrast, the parameter $a_6$ is determined by matching the speed of sound of previously defined intermediate polytrope.

The pressure as a function of energy density above $0.5\varepsilon_0$ is obtained by integration,
\begin{equation}
    P(\varepsilon) = P(0.5\varepsilon_0) + \int_{0.5\varepsilon_0}^{\varepsilon} \frac{c_s^2(\varepsilon')}{c^2}\, \mathrm{d}\varepsilon'.
\end{equation}

The free parameters
$(a_1, a_2, a_3, a_4, a_5, a_7)$
are assigned to the following distributions, which are determined based on the fundamental requirements for a physically viable EoS.
These requirements include achieving a reasonable maximum mass for neutron stars and ensuring adherence to the causal limit:
$a_1 \sim U(0.1,1.5)$,
$a_2 \sim U(1.5,12)$,
$a_3 \sim U(0.05a_2,2a_2)$,
$a_4 \sim U(1.5,37)$,
$a_5 \sim U(0.1,1)$,
$a_7 \sim U(0,1)$,
where $U(\cdot)$ denotes uniform distribution here.

In \cite{Greif2019}, one of the construction principles of this CS model is to approach the conformal limit at very high densities. However, since the conformal limit is expected to be satisfied by Perturbative Quantum Chromodynamics (pQCD) computations up to $50\,n_0$—and the densities in the neutron star interior are far below this threshold—in this study we relax the conformal limit condition commonly applied to the CS model. This relaxation is achieved by introducing $a_7$, a parameter that controls the  high density speed-of-sound limit in neutron star matter. The purpose of this adjustment is to explore a broader parameter space, potentially produce a more realistic mass range for twin stars, and investigate the substantial influence that the twin star condition may have on the speed of sound in the neutron star core. We discuss this motivation further in Appendix \ref{appendix}. Indeed, different modeling techniques have already suggested a non-conformal limit treatment by directly incorporating a constant speed of sound equal to the speed of light \citep{Zdunik_2013,Alford13,alford23}.

\begin{figure*}
    \centering
    \includegraphics[width=0.75\linewidth]{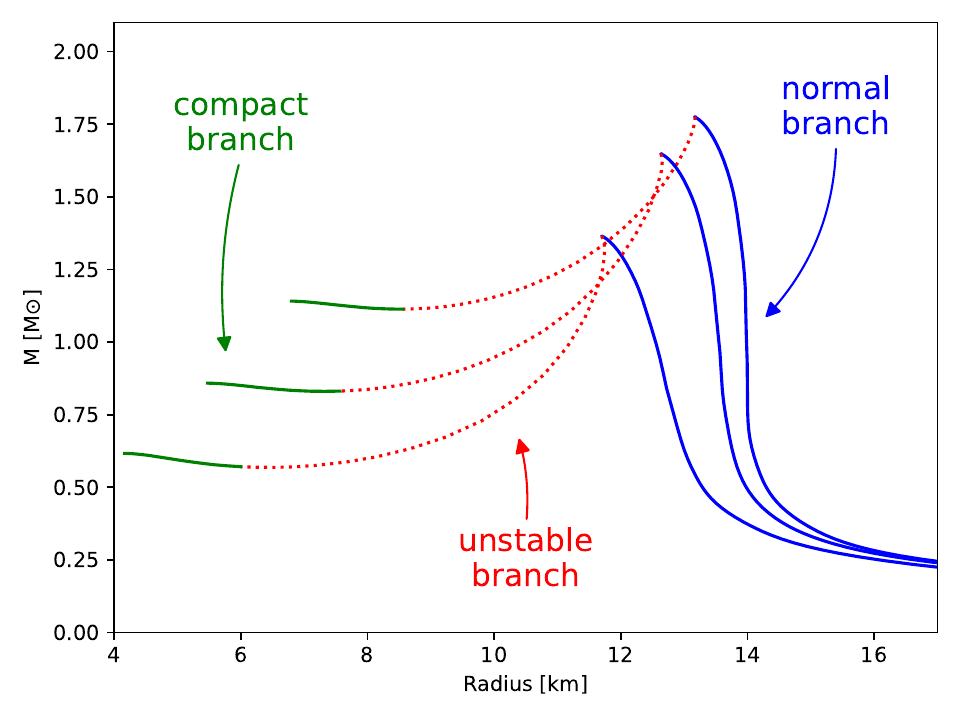}
\caption{
Illustration of three distinct MR curves, 
each fulfilling the constraint of the existence of a compact branch.
The curves depict mass (in 
\(M_{\odot}\)) as a function of radius (in km). 
As the central energy density increases, 
we initially encounter the blue solid line, 
which corresponds to the normal branch. 
Following this is the dotted red line, 
representing the unstable branch. 
Lastly, the green solid line indicates the compact branch. 
}
\label{fig:MR-curve-example}
\end{figure*}

The tension between approaching the conformal limit and accommodating the existence of neutron stars with mass exceeding 2.0$\msun$ has been extensively discussed in prior studies \citep{Bedaque:2014sqa,2013A&A...553A..22C,Alford13}. A clear relationship between the prior mass region and the speed-of-sound limit imposed in this CS model was also observed in our pre-study (see \hyperref[appendix]{Appendix} for more details). In light of this, we relax the conformal limit condition in this work and allow observational constraints to guide the determination of the appropriate speed-of-sound limit. This adjustment is intended to identify a limit that both supports the existence of twin star configurations and satisfies the constraints imposed by astrophysical measurements. 

To remain consistent with the original CS prescription, we impose the following constraints:
\begin{enumerate}
    \item Causality Constraint: The speed of sound, $c_s$, must lie within 0 and the speed of light, $c$.
    \item Fermi Liquid Theory Constraint: Up to $1.5\,\varepsilon_0$, the speed of sound should not exceed the limit suggested by Fermi liquid theory, i.e., $c_s^2(x=1.5)/c^2 < 0.163$ \citep{Greif2019,Schwenk2003,Schwenk2004}.
    \item Phase Transition Constraint: Since we are interested only in the EoS space that permits the occurrence of phase transitions, we impose a sufficient condition that $a_6 < 0$, as indicated by our formulation.
\end{enumerate}

The CS EoS employed in this study is continuous in the speed-of-sound domain, except within the phase transition region—hence, we refer to it as a `\textit{continuous}' EoS model. This feature distinguishes it from the polytropic (PP) model, which typically exhibits discontinuities in the speed of sound at the interfaces between different polytropic segments. In contrast, our chosen EoS represents a family of models that maintain a smooth speed-of-sound profile across the entire density range without abrupt changes.
This feature allows us to uniquely capture and explore the phase transition behavior inherent in such models. Our model is capable of undergoing phase transitions at any energy density and can accommodate a broad range of phase-transition energy depths.

Due to the functional form of the CS model defined in Equation \ref{cs_eq}—which is the sum of a logistic function and a Gaussian function—a phase transition cannot occur until the speed of sound reaches the maximum value defined by the Gaussian peak. Although this slightly limits our discussion, it remains effective given that this peak is a universal feature of viable EoS constructions, see \citet{Greif2019,Tews_2018} for further discussion. These considerations motivate our choice of EoS model.
\subsection{Twin-star definition} \label{sec:detect}
Due to the \textit{continuous} nature of the CS model, the vast majority of its parameter space produces EoS with \textit{smooth} behavior. However, for an EoS to support a twin star configuration, a strong phase transition must occur within the neutron star. In our study, the functional form (Equation \ref{cs_eq}) permits only first-order phase transitions. Thus, a strong first-order phase transition is a necessary condition for generating a twin star configuration. Nevertheless, a sufficiently strong phase transition alone does not guarantee the existence of twin stars. For this reason, our Bayesian inference framework incorporates an automated algorithm to detect twin star configurations. 
In this subsection, we will specifically discuss what we mean by twin star configuration.

For a given strong phase transition embedded EoS, we could solve The Tolman–Oppenheimer–Volkoff (TOV) equation to obtain the MR relation.
In Figure \ref{fig:MR-curve-example}, we illustrate several typical curves that all containing twin star configuration for demonstration. About twin star cases categrization, see discussion in \cite{Christian24,Montana:2018bkb}.
An MR curve that allows the existence of twin stars should consist of three segments:
As the central density begins to increase from a low value, 
we will observe a monotonically increasing mass with increasing of central density,
until the central density reaches the phase transition energy point.
This behavior is indicated by the solid blue line in Figure \ref{fig:MR-curve-example}. 
We define this branch as the `\textit{normal branch}' as it corresponds to stars composed entirely of normal matter before the onset of the phase transition.
When the central density exceeds the phase transition energy density, the MR curve undergoes a sharp turnaround: while the central density continues to increase, the total mass of the star decreases. This behavior characterized by $ {\mathrm{d}M}/{\mathrm{d}\varepsilon_c} < 0 $ indicates that this region is unstable (see Chap.16.3.9 of \cite{ferrari2020general}), as denoted by the dashed red line, we define this branch as `\textit{unstable branch}'.
Continuously increasing the central density leads to a subsequent increase in total mass of neutron star again, 
which corresponds to the twin-star branch we are investigating in this article, shown in the green solid line. Here, we define this branch as the `\textit{compact branch}' to emphasize that it comprises neutron stars with extreme compactness, highlighting that these hybrid twin stars are ultracompact in nature. 

In these specific given cases,
We note that a significant mass gap exists between the compact branch and the normal branch, spanning roughly from 7 to 12 km. The compact branch presented here has been investigated in various models (see, e.g., \citet{alford23}). However, in that study the speed-of-sound model is considerably stiffer than in our approach, which facilitates matching the twin stars over a wide range of MR space through parameter tuning. In contrast, due to the inherent characteristics of the CS model employed herein, the normal branch is predominantly governed by the Gaussian component of our CS construction. Consequently, we anticipate that constraints similar to those implemented in \cite{Raaijmakers_2021} and \cite{Rutherford_2024} will further restrict the behavior of this Gaussian part of this model. This constitutes the primary difference between our model and that of \cite{alford23}, who employ a constant speed-of-sound model rather than a Gaussian description. As discussed in \cite{Greif2019,Tews_2018}, we argue that the Gaussian speed of sound behavior for these normal matter represents a more reasonable choice.

\subsection{Bayesian inference framework}
\label{bayes_frame}
A Bayesian analysis has been conducted on this extended CS EoS model,
with the objective of exploring the EoS subspace that containing twin star configurations, 
and constraining this EoS subspace by current astrophysical observations. The framework described here follows the one developed by \citet{Greif2019} and \citet{Raaijmakers_2019,Raaijmakers_2020,Raaijmakers_2021}.
We denote all parameters of interest by $\bm{\theta}$, 
which contains the EoS parameters $(a_1,a_2,a_3,a_4,a_5,a_7)$ and the central energy density $\varepsilon_c$ of each MR measurement we are imposing.
The central density parameter is essential for representing the central densities of the MR measurement under this EoS framework. 
Bayes' theorem provides a framework to calculate their posterior distribution from the prior distribution $\pi$ and likelihood functions $\mathcal{L}$ defined by different measurements,
\begin{equation}
    \mathcal{P}(\bm{\theta} | \mathcal{D}) \propto 
    \pi(\bm{\theta}) \mathcal{L}(\mathcal{D} | \bm{\theta})
    ,
\end{equation}
where $\mathcal{D}$ denotes observation data, 
$\pi(\bm{\theta})$ is prior distribution of parameters, 
$\mathcal{L}(\mathcal{D} | \bm{\theta})$ is called likelihood function.
More specifically, the likelihood function can be rewritten as the product of the likelihood functions corresponding given dataset $\mathcal{D}$,
\begin{equation}
    \mathcal{L}(\mathcal{D} | \bm{\theta})\equiv
    \mathcal{L}(\mathcal{D} | \bm{M},\bm{R})\propto
    \mathcal{P}(\bm{M},\bm{R} | \mathcal{D})=
    \prod_{i=1}^{s} \mathcal{P}({M}_i,{R}_i | \mathcal{D}_i),
\end{equation}
where we use each set of model parameters $\bm{\theta}$ to generate $s$ observed stars.
From a particular set of parameter $\bm{\theta}$, 
we solve the Tolman–Oppenheimer–Volkoff (TOV) equations to get the 
corresponding mass $M_i$ and radius $R_i$.
The probability $\mathcal{P}({M}_i,{R}_i | \mathcal{D})$ is obtained by using the
kernel density estimate (KDE) technique from the observation data $\mathcal{D}$,
see the following section \ref{sec:data} for details.

For the prior parameter distribution $\pi(\bm{\theta})$, 
we use uniform distribution for model parameter $(a_1,a_2,a_3,a_4,a_5,a_7)$
within the given range in Section \ref{sec:EoS},
and the logarithm of central density follows uniform distribution that
$\log_{10}(\varepsilon_c / \mathrm{g\ cm^{-3}}) \sim U(14.7,17.0)$.
The reason we need such high central density upper limit is because we want complete the
compact branch MR space for given EoS setup. 
In our calculation, as shown in Figure \ref{fig:MR-curve-example}, 
the central density lower bound and upper bound 
for the compact branch (green line) is 
$\log_{10}(\varepsilon_c / \mathrm{g\ cm^{-3}}) \simeq 15.98^{+0.13}_{-0.17} \sim 16.23^{+0.26}_{-0.29}$. The lower bound is defined by the one of the turning point of the MR curve, marking the transition from the unstable branch to the compact branch. The upper bound is defined by the last stable point of the compact branch; beyond this point, further increases in central density result in a decrease in mass again. This is why it is called an `\textit{ultracompact twin star}' configuration. Such a high-density configuration has also been discussed in \cite{alford23}.

Note that some central densities yield MR points located on the unstable branch. In such cases, we assign an effectively negative infinite likelihood to effectively exclude these points from our analysis.
The posterior distribution of $\bm{\theta}$ therefore becomes
\begin{equation}
    \mathcal{P}(\bm{\theta} | \mathcal{D}) \propto 
    \pi(\bm{\theta}) \prod_{i=1}^{s} \mathcal{P}({M}_i,{R}_i | \mathcal{D}_i)
    .
    \label{eq:param_post}
\end{equation}
To sample the posterior distribution in Equation \ref{eq:param_post},
we use the nested sampling Monte Carlo algorithm
MLFriends \citep{Buchner_2014,Buchner_2019} with the \texttt{UltraNest}
package \citep{Buchner_2021}. All the inference carried out in this paper utilize the \texttt{CompactObject} package, an open-source full-scope neutron star equation of state inference pacakge developed by the authors \citep{EoS_inference,Huang:2024rfg}. Here, the construction of the CS model is kept consistent with another EoS inference package, which more focused on agnostic models, \texttt{NEoST} \citep{Raaijmakers2025}. This consistency ensures that our model construction aligns with previous studies like \citet{Raaijmakers_2019,Raaijmakers_2020,Raaijmakers_2021,Rutherford_2024}.

However, note that the prior range defined above does not represent the actual prior space considered in this work. In this study, we primarily focus on the strong first-order phase transition region of the CS model. More specifically, we perform inference on the model subspace that yields the existence of compact branches, which significantly diverges from the originally defined CS EoS parameter space (see Figure~\ref{fig:para-corn1}). Within this subspace, the resulting MR space naturally divides into two distinct branches: compact branch and normal branch. The region connecting these branches is unstable, as it exhibits decreasing compactness with increasing central density.

With the prior space appropriately defined, the following subsection describes the inference dataset utilized in this work.

\subsection{Inference Dataset}\label{sec:data}
One group of observations we employ comprises the MR measurements inferred from NICER data on three rotation-powered millisecond pulsars—PSR J0740+6620 (J0740), PSR J0030+0451 (J0030), and PSR J0437--4715 (J0437). Following this, we compare the resulting constraints with the MR measurement of the accretion-powered millisecond X-ray pulsar XTE J1814--338 (J1814). In this subsection, we clearly state the MR posterior files used in our analysis.

For PSR J0740+6620, a massive millisecond pulsar with a mass of approximately $2.0\,M_\odot$ \citep{Fonseca_2021}, the initial radius modeling was conducted by \citet{Riley2021,Miller2021}, and later revisited in \cite{Salmi24,Dittmann24}. Here, we use the MR posterior samples provided by \cite{Salmi24} as our standard input. This observation is particularly significant, as it is the most massive neutron star observed to date, and it is expected to provide a strong constraint on the EOS parameter space.

The pulsar PSR J0030+0451 is more challenging, having been initially analyzed in 2019 by different groups \citep{Riley2019,Miller2019} and subsequently reanalyzed in 2024 using a new background file \citep{Vinciguerra24}. This work seems demonstrated that the choice of different X-ray hotspot models can have a drastic influence on the MR measurement for PSR J0030+0451. Detailed modeling of the X-ray hotspot with a physics-motivated temperature distribution has been explored in \cite{huang2025physicsmotivatedmodelspulsar}. In this work, we employ the MR posterior for PSR J0030+0451 obtained with the ST+PDT hotspot model from \citet{Vinciguerra24}, as this model is considered more plausible and consistent with the magnetic field geometry inferred from the gamma-ray emission of the source.

For PSR J0437--4715, we use the MR posterior file corresponding to the hotspot model CST+PDT from \cite{Choudhury24}, which is based on a joint inference of XMM-Newton and NICER observational data.

Finally, we consider the RXTE observation of XTE J1814--338, recently analyzed by \cite{Kini_2024}, The MR posterior file we are using here is the \texttt{Bkg constrained} case. This source is of particular interest due to its unusually small mass and radius, which may indicate the presence of a twin star configuration and is consistent with the compact branch we are investigating.
\section{Result}\label{sec:result}
After establishing our method, we now present the results as follows. First, since our discussion is restricted to the twin star existence subspace of the CS model, we redefine the original (raw) prior space to a constrained prior that encompasses only the parameter space permitting twin star configurations. With this constrained prior, we derive the posterior using the MR constraints from current observations of the rotation-powered millisecond pulsars PSR J0740+6620, PSR J0030+0451, and PSR J0437--4715, all of which are expected to locate on the normal branch. Next, we computed the observational constraint specific to the compact branch by considering the MR measurement of XTE J1814--338. Finally, we combine all available observational constraints from both the compact branch (J1814) and the normal branch (J0740, J0030, and J0437) to impose an overall constraint on the twin star parameter space.
\begin{figure*}
\centering
\includegraphics[width=0.75\linewidth]{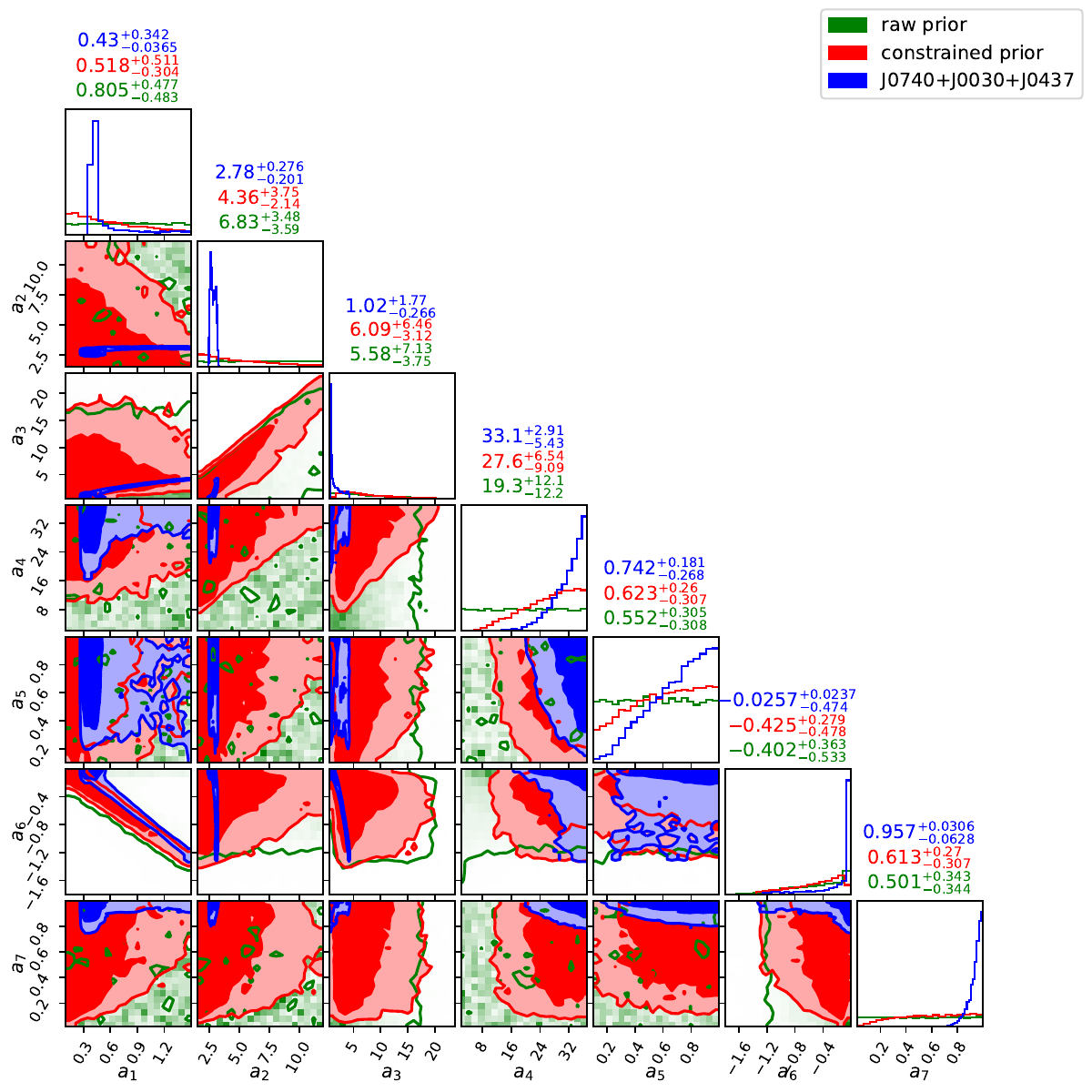}
\caption{
The posterior distributions of the seven EoS parameters $(a_1, \dots, a_7)$ under two distinct priors and the posterior constrained by MR measurements J0740, J0030 and J0437. 
The two different priors include the raw prior, which utilizes only the original CS model parameter space and is denoted by the green color, and the constrained prior that require the existence of a compact branch, indicated by the red color. 
The blue color represents the posterior after constraint by MR measurements J0740, J0030 and J0437. 
The two contour levels in the corner plot correspond to the 68\%, and 95\% credible intervals. 
The title of the 1-D marginal plot denotes both the median of the distribution and the range of the 68\% credible interval. 
}
\label{fig:para-corn1}
\end{figure*}

\subsection{Constrained prior}
The primary objective of this study is to investigate the under-explored parameter space associated with strong phase transitions that facilitate the existence of twin stars. As described in Section \ref{bayes_frame}, this parameter space defines the effective prior range, with the twin-star existence condition imposing the initial constraint. In Figure \ref{fig:para-corn1}, we compare the unconstrained prior range with the constrained prior range that permits twin star formation. Relative to the uniform priors defined for most EoS parameters, the twin star condition reduces the prior space to a smaller subspace.

However, the twin star constraint does not significantly refine the EoS parameters; rather, most of the EoS parameter space naturally yield twin star configurations. This observation suggests that twin stars are not rare phenomena but are intrinsically embedded within the majority of the model’s parameter space. Although direct observations of this compact branch remain elusive, its ubiquitous presence has been largely overlooked in many current studies employing the same model. The primary focus of this investigation is to explicitly demonstrate this twin star parameter subspace and constrain the EoS parameters using current observations, thereby providing deeper insights into the underlying twin star parameter space.

\begin{figure*}
\centering
\includegraphics[width=0.49\linewidth]{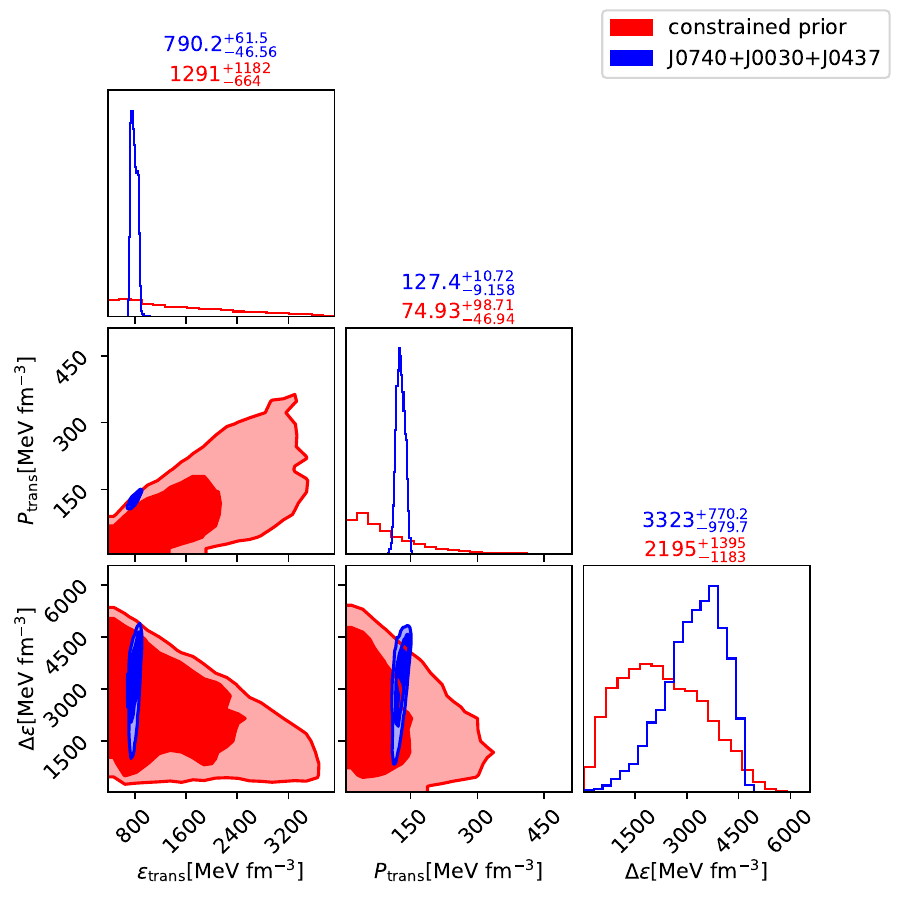}
\includegraphics[width=0.49\linewidth]{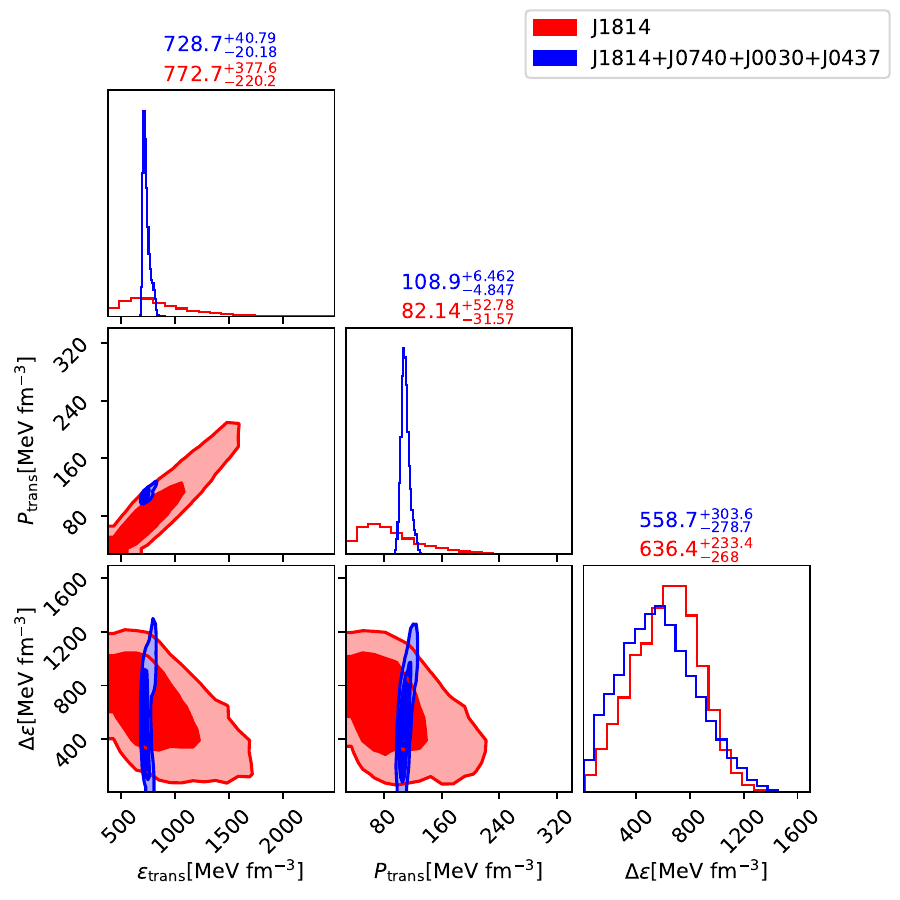}
\caption{
Left Panel: 
The corner plot illustrating the distribution of three phase transition parameters resulting from the EoS posteriors. 
The phase transition parameters under consideration include the transition point of energy density \( \varepsilon_{\text{trans}} \), the transition point of pressure \( P_{\text{trans}} \), and the transition energy depth \( \Delta \varepsilon \). 
The red color represents our constrained prior, while the blue color corresponds to constraint combination from MR measurements J0740, J0030 and J0437. 
The contour levels in the corner plot correspond to the 68\% and 95\% credible levels. 
The title of the 1-D marginal plot indicates both the median of the distribution and the range of the 68\% credible interval. 
Right Panel: 
This panel is analogous to the left panel, with the red color representing constraint from MR measurements J1814 and the blue color denoting constraint combination from MR measurements J0740, J0030, J0437 and J1814. 
}
\label{fig:pt-corn1}
\end{figure*}

In the left panel of Figure \ref{fig:pt-corn1}, we translate the twin star existence EoS parameter subspace into the phase transition parameter space. 
In our framework, the energy density and pressure at the critical phase transition point are determined through a root-finding procedure, which also facilitates the derivation of the transition depth.
Given a specific set of EoS parameters $(a_1, \dots, a_7)$, 
we can utilize a root-finding algorithm to compute the region 
$\varepsilon \in (\varepsilon_\text{trans},\varepsilon_\text{trans}+\Delta \varepsilon)$ 
where $c_s^2 < 0$ , according to Equation \ref{cs_eq}.
Then we can recognize the phase transition energy density as $\varepsilon_\text{trans}$,
retrieve the phase transition depth as $\Delta \varepsilon$, 
and compute phase transition pressure using
$P_{\text{trans}} = P(0.5\varepsilon_0) + \int_{0.5\varepsilon_0}^{\varepsilon_\text{trans}} {c_s^2(\varepsilon')}/{c^2}\, \mathrm{d}\varepsilon'.$

This transformation is beneficial as it directly relates the EoS parameters to phase transition quantities, emphasizing that only a sufficiently large phase transition can generate the twin star phase.  When comparing it to conventional phase transition model investigations (e.g., in \cite{huang2025,2025arXiv250115810L}), the phase transition depth automatically required by Bayesian inference is considerably smaller than the value proposed here, thereby revealing an unexplored parameter space even within Bayesian studies focusing on first-order phase transitions.

In Figure \ref{fig:mr-jg}, the constrained MR prior space, which satisfies the twin star existence condition, is outlined by black and green dotted lines. Due to the intrinsic discontinuity between the compact and normal branches, the contour is presented as two distinct regions rather than as a continuous weighting of both branches. Although these regions exhibit some overlap, within any single MR curve, the compact and normal branches do not intersect due to the intervening unstable region. This overlap suggests that, under certain finely tuned EoS parameter choices, the MR space corresponding to one branch may extend into that of the other. Additionally, only a small fraction of the 99\% MR space for the compact branch exceeds the 1.0 $\msun$ limit, which is commonly regarded as the lower mass threshold for neutron stars based on formation theories.

This twin-star-restricted parameter space constitutes our constrained prior space. In the subsequent section, we use this space as a reference for comparison, to constrain the EoS parameter subspace responsible for producing twin stars. 
Focusing exclusively on the regions within a continuous model (such as the CS model implemented here) that can accommodate twin stars, and exploring the possibility of directly imposing astrophysical constraints on this subspace, is of importance.

\subsection{Current observation constraints}
After defining our constrained prior space in detail, we now proceed with the constraints of interest. In Figure \ref{fig:para-corn1}, we present the posterior distribution after applying constraints from all MR measurements extracted from NICER data (J0740, J0030, and J0437). For most of the parameter space, the constraints narrow the allowed region considerably. A particularly noteworthy observation is related to the parameter $a_7$, which represents the asymptotic value of the speed of sound squared at high density. This parameter clusters around a value of 1, indicating that the speed of sound approaches the speed of light within stellar core. While the conformal limit of $1/3$ lies well outside our 3-$\sigma$ credible interval, this result may implies the presence of more exotic degrees of freedom, such as deconfined quark matter \citep{cslimit_steiner,Tews_2018}. 

\begin{figure*}
    \centering
    \includegraphics[width=0.8\linewidth]{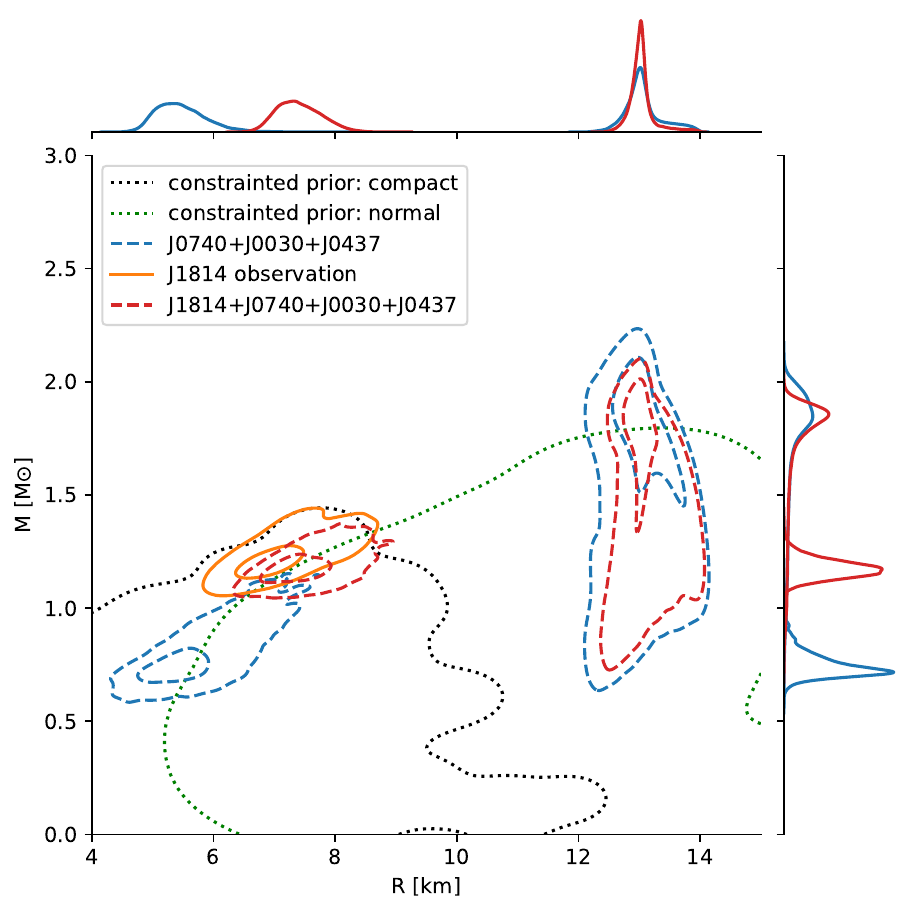}
    \caption{
MR contour plot illustrating three distinct constraints alongside one astronomical observation.
The orange solid line signifies the 68\% and 99\% credible levels for star J1814's MR measurement directly from observation. The dotted line represents the 99\% credible level of our constrained prior space, where the black and green colors denote the compact branch and normal branch, respectively.
The blue dashed lines indicate the 68\% and 99\% credible levels of posteiors for the constraint combinations from MR measurements J0740, J0030 and J0437. The red dashed indicate the 68\% and 99\% credible levels of posteriors for the constraint combinations from MR measurements J1814, J0740, J0030 and J0437. The compact branch and normal branch region plot separately.
}
    \label{fig:mr-jg}
\end{figure*}

As we did before, we can readily translate these EoS posterior to the phase transition parameter space, as illustrated by the blue contours in the left panel of Figure \ref{fig:pt-corn1}. The phase transition parameters offer a more physically interpretable framework. It is evident that the phase transition point is strongly confined to a narrow peak in the $\varepsilon-P$ plane, with an order-of-magnitude reduction in uncertainty. This behavior is consistent with the constraints imposed by NICER MR measurements—particularly for J0740, a 2.0 $\msun$ neutron star—which has drastically narrowed the allowed region for the phase transition point to a narrow peak around the 2.0 $M_{\odot}$ threshold. To satisfy the 2.0 $\msun$ maximum mass condition, the phase transition point is positioned at a significantly larger value compared to previous studies (e.g., \cite{huang2025}) to reach this limit before phase transition happened. It should be noted that this outcome is partly a consequence of the inherent characteristics of the employed EoS model, which requires the maximum mass condition to be met before the occurrence of the phase transition, with the normal matter represented solely by a Gaussian peak in the speed of sound.

In contrast, the constraint on the phase transition depth, $\Delta\varepsilon$, is comparatively weaker. Nevertheless, $\Delta\varepsilon$ is constrained to a high value of approximately $3323_{-980}^{+770}$ MeV/fm$^{-3}$. This indicates that, even though the current MR constraints primarily target normal neutron star matter, the observations already allow us to refine the twin star phase transition space—particularly by tightly constraining the transition point. However, due to the limited information regarding ultra-compact twin stars, the phase transition depth remains substantially less well-constrained.

In Figure \ref{fig:mr-jg}, we present the MR posterior contour derived from this EoS posterior. The normal branch, outlined by the blue dashed line, encompasses all the observed sources (J0740, J0030, and J0437), yielding a posterior distribution characterized by 
$
M = 1.8^{+0.12}_{-0.43}\ M_\odot \quad \text{and} \quad R = 13.0^{+0.34}_{-0.21}\ \text{km}
$
(see Table \ref{tab:mr} for details). This MR region is comparable to, yet more tightly constrained than, those reported in \cite{Raaijmakers_2021,Rutherford_2024}. Since only the twin star constrained prior is explored in this regime.

The compact branch features stars with exceptionally high central densities—exceeding $10^{16}$ g/cm$^3$—but with masses predominantly below 1.0 $M_\odot$ and radii smaller than 8 km. This finding in tension with the conventional expectation that neutron stars typically have masses exceeding 1.0 $M_\odot$. Although certain quark star models could, in theory, reproduce such MR characteristics, these models usually predict significantly lower central densities. Instead, our model may indicate the existence of extremely dense, low-mass, yet compact neutron stars.

Recently, as noted in the sections above, a new MR measurement of J1814 was reported by \citet{Kini_2024}. This neutron star is found to lie approximately within the MR compact branch posterior region, although its central mass and radius are slightly larger than those of the primary compact branch posterior region (see Figure \ref{fig:mr-jg}). This direct overlap is interesting because it suggests that, even without incorporating additional constraints, the compact branch emerging from this model has the potential to account for this unusual MR measurement. Consequently, J1814 might be interpreted as a twin star residing on the compact branch rather than on the normal branch.
\begin{figure*}
    \centering
    \includegraphics[width=0.49\linewidth]{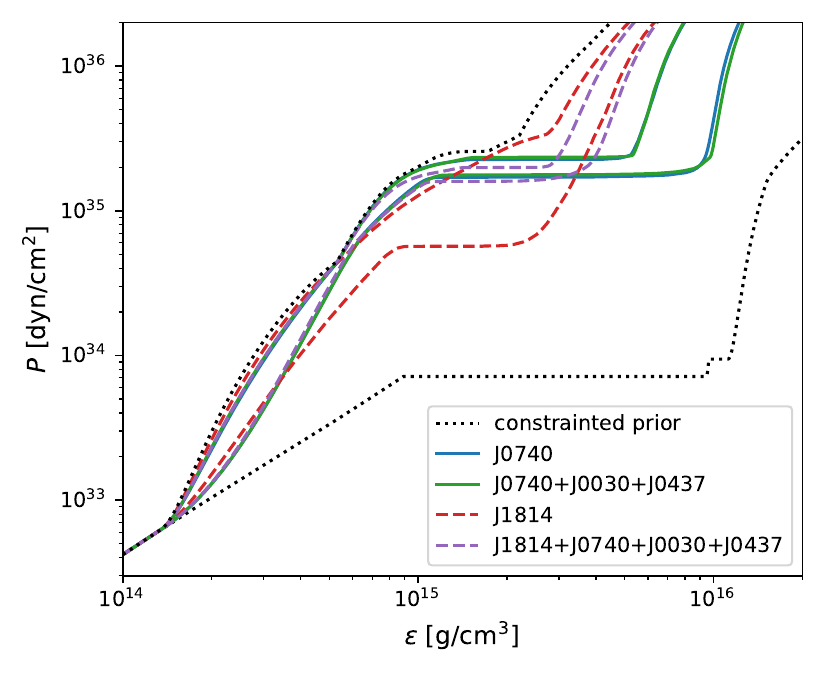}
    \includegraphics[width=0.49\linewidth]{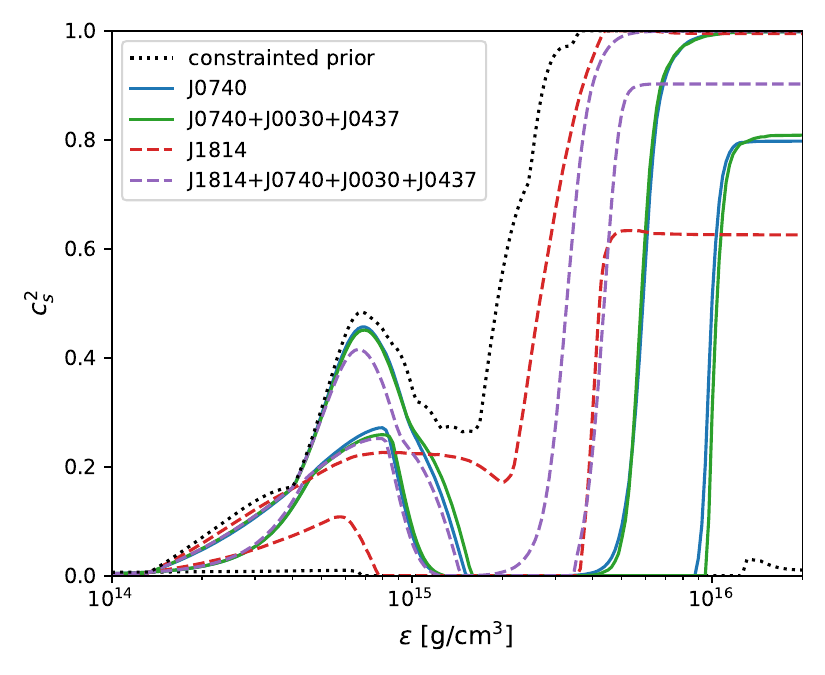}
\caption{
Left Panel: 
The 95\% credible level band for the EoS in the \(\varepsilon-P\) plane, illustrating one constrained prior and four posteriors. 
The constrained prior is represented by a black dotted line. 
The different set of posteriors are depicted as follows: the blue solid line for constraint from J0740, the green solid line for constraint from J0740, J0030 and J0437, the red dashed line for constraint from MR anomaly J1814, and the purple dashed line for posterior constrained by all availiable MR measurements J0740, J0030, J0437 and J1814. 
Right Panel: 
The 95\% credible level band for the squared sound speed \(c_s^2\) in the \(\varepsilon-c_s^2\) plane. 
All color, line style, prior, and posterior correspondences are consistent with those depicted in the left panel, differing only in the y-axis label. 
}
\label{fig:all-EoS-cs}
\end{figure*}
\subsection{Direct Constraint on Compact Branch from XTE J1814--338}
\begin{table*}
\begin{tabular}{|c|c|c|c|c|}
\hline
& \multicolumn{2}{c|}{Compact branch} & \multicolumn{2}{c|}{Normal branch} \\ \hline 
Constraint Dataset & $M\ [M_\odot]$ & $R\ [\rm km]$ & $M\ [M_\odot]$ & $R\ [\rm km]$ \\ \hline
J0740 &$0.736^{+0.069}_{-0.033}$ & $5.403^{+0.448}_{-0.360}$ & $1.852^{+0.100}_{-0.500}$ & $12.988^{+0.116}_{-0.190}$              \\ \hline 
J0740+J0030+J0437 &$0.735^{+0.069}_{-0.033}$ & $5.420^{+0.449}_{-0.354}$ & $1.807^{+0.120}_{-0.434}$ & $13.022^{+0.341}_{-0.208}$              \\ \hline 
J1814 &$1.156^{+0.058}_{-0.076}$ & $7.192^{+0.916}_{-0.564}$ & $1.252^{+0.164}_{-0.197}$ & $12.838^{+0.846}_{-1.142}$              \\ \hline
J1814+J0740+J0030+J0437 &$1.172^{+0.044}_{-0.038}$ & $7.375^{+0.408}_{-0.333}$ & $1.793^{+0.080}_{-0.419}$ & $13.017^{+0.124}_{-0.132}$              \\ \hline 
\end{tabular}
\caption{ The median and 68\% credible intervals for the MR posterior marginal distributions of the normal branch and compact branch on different constraint dataset combinations. }
\label{tab:mr}
\end{table*}

As explained above, incorporating MR measurements from NICER on the normal branch of neutron stars already naturally yields a twin star space that aligns with the MR measurement reported for XTE J1814--338. This correspondence may be interpreted as an indication of the existence of compact branches. However, due to the limited observational constraints on compact branch, the distribution of the phase transition depth $\Delta\varepsilon$ remains broad, with much of it having been excluded by previous studies eg. \citet{huang2025}.

Despite a qualitative consistency between the MR posterior range for the compact branch and the MR measurement of J1814, we can also perform Bayesian inference using J1814 directly as a constraint to evaluate its impact on constraining the compact branch MR space. This approach enables a quantitative assessment of how well this observation aligns with the tentative twin star EoS parameter space.

As shown in the right panel of Figure \ref{fig:pt-corn1}, the red contour represents the constraint on phase transition parameters based solely on the J1814 observation. This constraint substantially narrows the allowed range for the phase transition depth, $\Delta\varepsilon$, compared to the left panel. This is expected because the J1814 observation directly probes the compact branch, thereby determining the range of the phase transition region. The posterior of phase transition depth is $636.4_{-268}^{+233}$ MeV/fm$^3$.

These findings offer a clearer picture of the impact of different observational constraints on the EoS: MR measurements of J0740, J0030, and J0437 primarily constrain the phase transition point on the normal branch, constraints from compact-branch observations like J1814 effectively limit the phase transition depth.

Combining all the constraints from these two groups of MR measurements yields a more comprehensive restriction on the phase transition parameters—both the transition point and the transition depth ($\Delta\varepsilon$). This is illustrated by the blue contour in the right panel of Figure \ref{fig:pt-corn1}. Both the posterior of phase transition point and the transition depth have been constrained to a reasonable range. In \cite{Christian24}, the authors investigated several phase transition scenarios and provided a comprehensive roadmap outlining which phase transitions are viable under multiple constraints. With the combination of astrophysical constraints and the twin star existence condition, our EoS exploration based on this CS model yields a parameter space that not only accommodates twin stars but also fully satisfies the astrophysical constraints. Moreover, compared to the parameter space diagram in 
\cite{Christian24}, our inferred posterior range—particularly its central value—lies far outside the conventional astrophysical observation allowed region, yet remains within the twin star–allowed space.

From this inference, 
we can extract the central density of J1814 based on this given EoS model.
For the constraint implemented by only J1814 constraint on compact branch, 
the posterior central density of J1814 is
$\log_{10}(\varepsilon / \mathrm{g\ cm^{-3}})=15.855^{+0.058}_{-0.069}$
and for constraint introduced by whole NICER observations plus J1814. 
We have the posterior central density equals to
$\log_{10}(\varepsilon / \mathrm{g\ cm^{-3}})=15.895^{+0.035}_{-0.042}$ . 
This is indeed an ultra-dense configuration, with very large central density.

Projecting these posterior distributions onto the MR space provides further insight into the astrophysical constraints. As shown in Figure \ref{fig:mr-jg}, the red dashed contours represent the combined constraints from J0030, J0740, J0437, and J1814. When comparing the normal branch with and without the J1814 constraint, the maximum mass of the region decreases from approximately 2.5 $M_\odot$ to 2.2 $M_\odot$, and the radius distribution becomes correspondingly narrower. This is an interesting phenomenon: the astrophysical constraint on the compact branch (with and without J1814) could influence the normal branch parameter space. In the compact branch, the MR posterior distribution nearly completely overlaps with the input MR measurement of J1814 (as indicated by the orange solid contour line), thereby underscoring the capability of the compact branch to account for this observation. The proposed twin star existence range in this branch is $1.172^{+0.044}_{-0.038}\ M_\odot$. Consequently, the J1814 constraint demonstrates that the CS model is capable of producing a reasonable twin star configuration.

In \cite{Kini_2024}, the authors acknowledge significant systematic uncertainties in the J1814 modeling results, noting that the single hotspot computation may not be entirely reliable. They also suggest that the observed small mass and extremely small radius could imply a phase transition pressure below 50 MeV/fm$^{-3}$. In contrast, our analysis shows that, under the assumption of twin stars, the observed MR data can be reconciled with a reasonable phase transition pressure. By incorporating current astrophysical observations, we infer a phase transition pressure of $108.9_{-4.847}^{+6.462}$ MeV/fm$^{-3}$, which is considerably higher than 50 MeV/fm$^{-3}$.
\subsection{$\varepsilon$-$P$ and Speed of sound posterior}
Since this model is intrinsically generated by modulating the speed of sound in neutron star matter, directly plotting the resulting EoS and speed of sound posteriors is particularly insightful. In Figure \ref{fig:all-EoS-cs}, we show the posteriors that incorporate all the constraints implemented. Notably, the initially extensive EoS space that permits twin star configurations is refined into a narrow region. For astrophysical constraints excluding J1814 (J0740 and J0740 + J0030 + J0437), the posterior exhibits a broad plateau in the intermediate EoS region, suggesting that the underlying EoS encompasses a large phase transition. The inclusion of the J1814 constraint further refines the EoS space by constraining the phase-transition depth to a narrower range.

Moreover, the posteriors obtained from analyses based solely on J0740, as well as combinations of J0740 with J0030 and J0437, reveal a highly overlapping EoS space, which corresponds to a similar MR region. This consistency explains why gravitational wave constraints from GW170817 \citep{LIGOScientific:2017vwq,Abbott_2018} were not included in this inference, since the component stars of GW170817 locates in a mass range similar to those of J0437 and J0030, and thus offer no significant additional constraining power beyond that provided by J0740.

In the right panel of Figure \ref{fig:all-EoS-cs}, the posterior is projected onto the speed of sound parameter space.  Under the twin star constraint alone, the speed of sound in the high-density region is surprisingly not tightly constrained. The 95\% credible contour reveals that a substantial portion of the parameter space remains permissible, which is contrary to our expectation that a stiff EoS would be required to generate twin stars. However, when only the J0740 constraint is applied, the high-density behavior of the speed of sound is confined to a region where 
$
c_s^2/c^2 > 0.8.
$ When all constraints are applied, we confirm that the high-density speed of sound is bounded from below by $ c_s^2/c^2 > 0.9 $. This result reinforces our earlier discussion: the combination of the twin star condition and the astrophysical constraints necessitates that the speed of sound in neutron stars approaches the speed of light rather than conforming to the conformal limit, as also noted in \cite{Alford13,cslimit_steiner, Tews_2018}. Such a high, nearly constant speed of sound suggests that the neutron star core in this branch is likely to contain exotic degrees of freedom.
\section{Conclusion and Discussion}\label{sec:discuss}
In summary, we have systematically explored—for the first time—the parameter space within a widely adopted speed-of-sound (CS) EoS meta-model to investigate the emergence of twin star configurations via strong first-order phase transitions. This subspace is surprisingly broad, supplementing existing studies that have not accounted for it. Our comprehensive Bayesian analysis—integrating MR constraints from NICER observations of rotation-powered millisecond pulsars like J0030, J0740, J0437—reveals that the resultant compact branch, situated around 1 to 1.2 $M_{\odot}$ with a radius of approximately 7 km, surprisingly coincides with the MR ranges proposed for the anomalous observation of XTE J1814--338. This concordance suggests that hybrid twin star configurations may provide a viable explanation for the observed anomaly.

Futhermore, although \citet{Kini_2024} propose that accommodating such an ultracompact neutron star might require a phase transition pressure $P_{\text{trans}} < 50$ MeV/fm$^3$, our compact branch yields an extreme transition pressure of $P_{\text{trans}} = 108.9_{-4.85}^{+6.46}$ MeV/fm$^3$, transition density $\varepsilon_{\text{trans}}/\varepsilon_0 = 4.847_{-0.134}^{+0.271}$ and an energy density jump $\Delta \varepsilon = 636.4_{-268}^{+233}$ MeV/fm$^3$, corresponding to $\Delta \varepsilon/\varepsilon_0 = 3.716_{-1.854}^{+2.020}$. By analyzing the speed of sound behavior of the EoS posterior, we found the high-density speed of sound is driven toward the speed of light ($c_s^2/c^2 > 0.9$) to full fill all the astrophysical constraints if assume the existence of twin star, indicating the potential presence of some exotic matter in the hybrid twin star core. Overall, our findings demonstrate that continuous CS meta-model naturally accommodate both conventional and ultra-compact neutron star configurations, offering new insights into the exotic physics governing dense matter.

In this discussion, we did not take into account the observation of HESS J1731-347. Recent observations \citep{Doroshenko22} suggest that this object could represent a measurement in which its MR fall within the radius gap between our compact and normal branches. However, as noted in \cite{Salmi:2024bss}, significant systematics remain in these results, such as the distance assumptions based on a uniform-temperature carbon atmosphere model (see also \cite{alford23}). Considering these caveats, we have chosen not to include this object in our analysis of twin star cases.

Another notable source missing from our current constraints is PSR J1231-1411 \citep{Salmi:2024bss,Qi_2025}, a millisecond X-ray pulsar reported to have a mass of $1.04_{-0.03}^{+0.05}\,M_{\odot}$ and a radius of $13.5_{-0.5}^{+0.3}\,\mathrm{km}$. Although there remain considerable systematic uncertainties related to inference convergence and modeling strategies about this source, it is interesting to note that this source lies on the normal branch of our twin star MR posterior, indicating that PSR J1231-1411 would reside in a normal branch side of twin star space characterized by similar masses as we proposed for the compact branch but distinct radii, with a radius difference $\Delta R$ of approximately 6~km.

\section*{Acknowledgements}
H.C. extends gratitude to Mark Alford, Alexander Chen and Ryan O'Connor for insightful discussion. 
All the reproduction files and necessary datasets are available in Zenodo \citep{Zhou2025}.


\appendix 
\section{Relation between Prior mass range and speed of sound limit}
\label{appendix}
\begin{figure}
    \centering
    \includegraphics[width=0.75\linewidth]{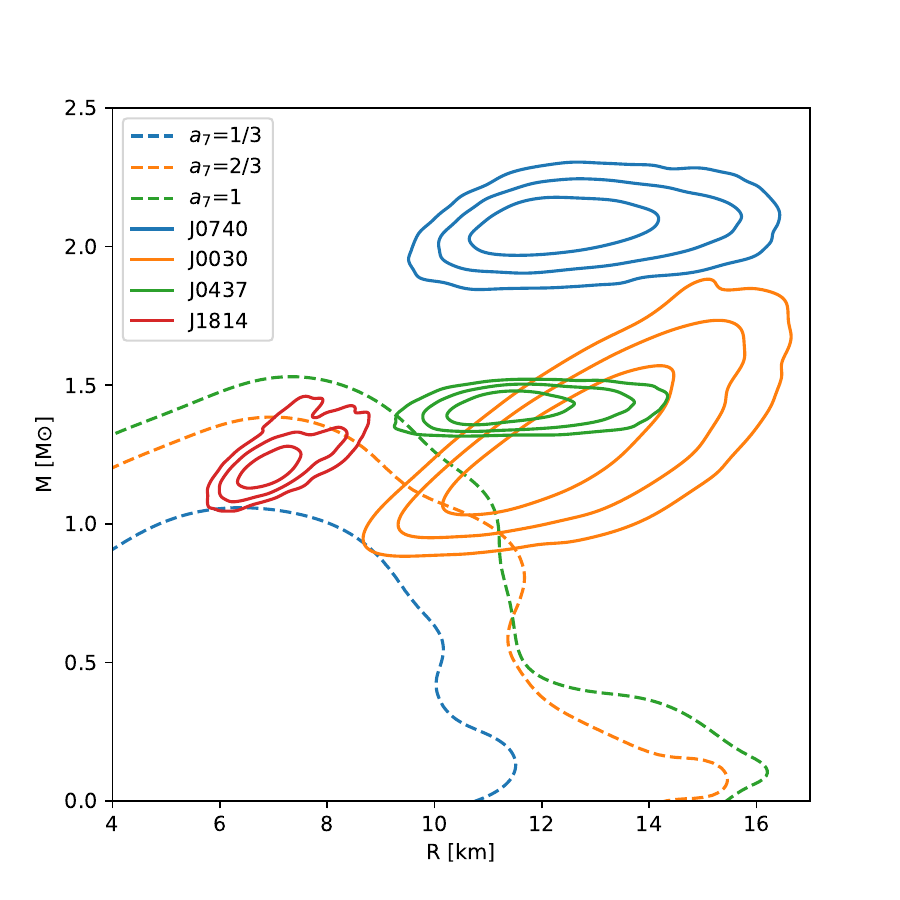}
    \caption{
    Contour plot illustrating the various MR distributions. 
    The solid lines represent the 68\%, 95\%, and 99\% credible levels derived from observations of four stars (J0740, J0030, J0437, J1814), with distinct colors corresponding to each star. 
    The dashed line indicates the compact-branch 99\% credible level MR region for varying values of \( a_7 \) within our constrained prior, with different colors denoting each values of \( a_7 = \{1/3,2/3,1\} \). 
}
    \label{fig:left-mr-a7}
\end{figure}

In Figure \ref{fig:left-mr-a7}, we show a clear relationship between the asymptotic limit of the speed of sound at very high densities and the MR range for the twin star configurations. In this plot, we show all points within the 99\% contour. Specifically, the MR range increases as the upper limit of the speed of sound is raised. For comparison, we also plot MR posteriors for four pulsars: PSR J0030+0451 (J0030, \cite{Riley2019,Miller2019,Vinciguerra24}), PSR J0437--4715 (J0437, \cite{Choudhury24}), PSR J0740+6620 (J0740, \cite{Riley2021,Miller2021,Salmi24,Dittmann24}), and XTE J1814--338 (J1814, \cite{Kini_2024}). Notably, Figure \ref{fig:left-mr-a7} shows that J1814 falls completely outside the constrained MR prior region corresponding to $a_7 = 1/3$, where $a_7$ is defined as the high-density limit of the speed of sound squared in this CS model. 
\bibliography{sample631}{}

\bibliographystyle{aasjournal}

\end{document}